\documentstyle [epsfig]{mn}
 at 10truept

\title
     [Photodissociation of $\rm{H}_{2}$]
{\vglue-3.0truecm
\vglue 2.5truecm
      On the photodissociation of $\bmath{\rm{H}_{2}}$ by the first stars
\author
     [S.C.O. Glover \& P.W.J.L. Brand]
     {S.C.O. Glover \& P.W.J.L. Brand\\
     Institute for Astronomy, 
     University of Edinburgh,
     Royal Observatory,
     Blackford Hill, 
     Edinburgh, 
     EH9 3HJ}}

\ifCUPmtlplainloaded \else
 
\fi

\newcommand{\Hm}{\rm{H}^{-}}
\newcommand{\Hp}{\rm{H}^{+}}
\newcommand{\me}{\rm{e}^{-}}
\newcommand{\mH}{\rm{H}}
\newcommand{\mHt}{\rm{H}_{2}}
\newcommand{\mHtp}{\rm{H}_{2}^{+}}

\begin{document}

\maketitle

\begin{abstract}     
The first star formation in the Universe is expected to take place within small
protogalaxies, in which the gas is cooled by molecular hydrogen. However, 
if massive stars form within these protogalaxies, they may suppress further 
star formation by photodissociating the $\mHt$. We 
examine the importance of this effect by estimating the timescale on which
significant $\mHt$ is destroyed. We show that photodissociation is significant
 in the least massive protogalaxies, but becomes less so as the protogalactic 
mass increases.
We also examine the effects of photodissociation on dense clumps of gas within 
the protogalaxy. We find that while collapse will be inhibited in low density 
clumps, denser ones may survive to form stars.
\end{abstract}


\section{Introduction}
Although we can now detect star-forming galaxies at very high redshifts,
the ionization and ubiquitous metal enrichment of the IGM at these redshifts
suggest that we are not yet seeing the very earliest star formation. In order 
to study the formation of the first stars, we are thus forced to 
rely on a purely theoretical approach. Work over the past few decades has led
to substantial progress in our understanding of this epoch, and while many of
the details are unclear, a broad outline of events can be constructed.

The first step is the gravitational collapse of overdense regions in the 
early universe. In cold dark matter models, this happens hierarchically, with
objects on the smallest scales forming first.
In order for gas to collapse along with the dark matter, however,
gravitational forces must be strong enough to overcome the thermal pressure
of the gas. This occurs for overdensities more massive than the 
Jeans mass, $M_{\rm J}$, which, prior to reionization, is of the order of 
$10^{4} \: M_{\odot}$~\cite{har}. 

As the gas collapses, it is heated by adiabatic compression and by shocks.
In the absence of cooling, this leads to an increase in pressure which
will eventually halt the collapse. Moreover, it prevents fragmentation of the 
gas. In order for star formation to occur, continued collapse and
fragmentation are required, and thus stars will form only if the gas can 
lose heat
effectively. In quantitative terms, it is necessary that the cooling timescale 
be shorter than the free-fall timescale \cite{ro}. 

During the collapse of the first protogalaxies, the gas temperature typically 
reaches only $1000$ -- $2000 \: \rm{K}$~\cite{teg}. This is significantly 
lower than the temperature
required for the gas to be able to cool via Lyman-$\alpha$ emission. 
In the absence of metals, the only effective low temperature coolant is 
molecular hydrogen.
This forms mainly via the reactions
\begin{eqnarray}
 \mH + \me & \rightarrow & \Hm + \gamma \label{hm} \\
 \Hm + \mH & \rightarrow & \mHt + \me, 
\end{eqnarray}
although a small amount also forms via
\begin{eqnarray}
 \mH + \Hp & \rightarrow & \mHtp + \gamma \\
 \mHtp + \mH & \rightarrow & \mHt + \Hp. 
\end{eqnarray}
While the final $\mHt$ fraction is generally small, it is nonetheless able
to cool primordial gas effectively, as has been established by a number of 
authors (see Haiman, Thoul \& Loeb 1996  and references therein).

The subsequent evolution of the gas and the eventual formation of stars 
are still not well 
understood, although substantial progress in this area has recently been made
(Abel, Bryan \& Norman 2000; 
Bromm, Coppi \& Larson 1999; Nakamura \& Umemura 1999).
 
Finally, once star formation begins, it will influence the surrounding gas via
a number of feedback processes. Two processes in particular have been 
studied: energy input from supernovae and UV radiation from stars.

Since the energy of a single supernova is comparable to the binding energy of 
the first protogalaxies, it is highly probable that they will
be entirely disrupted by supernovae (MacLow \& Ferrara 1999; 
Ferrara \& Tolstoy 2000).
 However, as this occurs only once 
the first massive stars reach the end of their life, it is ineffective 
during the first few million years of star formation. Moreover, the effects of
supernovae are quite local, being restricted to a single protogalaxy and its 
immediate surroundings.

Ultraviolet radiation, on the other hand, can potentially act much faster, and
over a much greater volume. It exerts a feedback by photodissociating $\mHt$, 
thereby suppressing cooling.
 The global effects of ultraviolet radiation 
from the first stars have recently been studied by Haiman et 
al.~\shortcite{har} and 
Ciardi et~al.\ \shortcite{cfgj}. They show that a soft ultraviolet background 
builds up prior
to reionization, which photodissociates $\mHt$ in the IGM and in newly 
collapsing protogalaxies.

Less work has been done on the local effects of UV radiation, ie.\ 
the ability of the first massive stars to suppress $\mHt$ cooling within their 
host protogalaxy. This has recently been studied by Omukai \& 
Nishi~\shortcite{on}, who find that
this local feedback is very effective, with the radiation from a single 
massive star sufficient to dissociate $\mHt$ throughout a 
protogalaxy. However, their treatment of the effects of the 
self-shielding of $\mHt$ is correct only once the photodissociation region 
created by the star has reached equilibrium. Without knowing the  
timescale on which this occurs, it is impossible to assess the true 
effectiveness of the local feedback from UV radiation.

In this paper we overcome this problem by introducing a simple method to 
estimate the photodissociation timescale. We first calculate the rate at
which photodissociating photons are produced by a primordial star, 
approximated as a blackbody. Next, we calculate how many of these photons 
illuminate the object of interest, how many are absorbed, and 
finally, how many of these absorptions lead to photodissociation.
Most of these quantities can be estimated to within a factor of a few or 
better; even in the worst case, we expect our final estimate to be accurate
 to within an order of magnitude.
 
Applying this method to simple protogalactic models, we find that the 
photodissociation timescale depends strongly upon the mass of the protogalaxy.
In the least massive protogalaxies, equilibrium is reached on a timescale that
is short compared to the stellar lifetime, while in larger protogalaxies, the 
two timescales are comparable. We also examine the fate of $\mHt$ in dense 
clumps of gas within the protogalaxy. Such clumps are the likely precursors of
 star formation, but can collapse only if enough $\mHt$ survives within them 
to provide efficient cooling. We find that this is only the case within the 
densest clumps.

We thus partially confirm the results of Omukai \& Nishi. Ultraviolet feedback 
is very effective at suppressing cooling in small protogalaxies, if the
amount of gas in dense clumps is small. It is less effective in larger systems,
or when much of the gas is densely clumped. This suggests that UV feedback 
alone need not imply a low star formation efficiency.

In section 2 we examine the manner in which UV radiation leads to the 
photodissociation of $\mHt$, and briefly discuss the approach taken by
Omukai \& Nishi and the importance of a proper treatment of self-shielding.
In section 3 we introduce our approximate model, and in section 4 apply it 
to our protogalactic model. In section 5 we examine the effects of a number
of complications not included in our simple model, and we conclude in section
6.

\section{Photodissociation of \protect{$\bmath{\rm{H}_{2}}$}}
The first generation of protogalaxies which are able to cool effectively via
$\mHt$ emission have virial temperatures of the order of $1000 \:
 \rm{K}$, and mean densities of the order of $1 \: \rm{cm}^{-3}$ \cite{teg}.
Under these conditions, the amount of $\mHt$ in an excited energy state will 
be very small, and it is a good approximation to consider all of the $\mHt$ as
being in the ground state (in either the ortho or para form, as appropriate).
In order to destroy ground-state $\mHt$ molecules, a substantial energy input 
is required. At higher temperatures ($T \geq 6000 \: \rm{K}$), this could be 
provided by collisions, but at $1000 \: \rm{K}$ collisional 
dissociation is unimportant. In this case, radiative reactions will dominate 
the destruction of $\mHt$. There are several possibilities.
\begin{enumerate}

\item Photoionization: 
\begin{equation}
 \mHt + \gamma \rightarrow \mHtp + \me
\end{equation}
\cite{or}. This has a threshold of $15.42 \: \rm{eV}$, well above the energy 
required for \hbox{H\,{\sc i}} ionization. It will only occur within the 
\hbox{H\,{\sc ii}} region surrounding the radiation source. As we show in 
section~\ref{hii}, this region may be quite small.

\item Excitation to the vibrational continuum of an excited 
electronic state of $\mHt$: 
\begin{equation}
\mHt + \gamma \rightarrow 2\mH  
\end{equation}
\cite{ad}. 
This has a threshold of $14.16 \: \rm{eV}$ for ortho-hydrogen and $14.68 \:
\rm{eV}$ for para-hydrogen. This too is restricted to \hbox{H\,{\sc ii}} 
regions.

\item Direct excitation to the vibrational continuum of the
ground electronic state of $\mHt$. This process is strongly forbidden
and proceeds at a negligible rate.

\item Two-step photodissociation (the Solomon 
process): 
\begin{equation}
 \mHt + \gamma \rightarrow \mHt^{*} \rightarrow 2\mH 
\end{equation}  
\cite{sw}. The first step involves excitation to either the 
$B^{1}\Sigma^{+}_{u}$ or $C^{1}\Pi_{u}$ 
excited electronic states, with threshold energies of $11.15 \: \rm{eV}$ 
and $12.26 \: \rm{eV}$
 respectively. This is followed by decay to some vibrational level of
the ground electronic state. A fraction of these decays will be to the 
vibrational 
continuum of the ground state, resulting in dissociation. A number of 
rotational and vibrational states of $B^{1}\Sigma^{+}_{u}$ and $C^{1}\Pi_{u}$ 
 are available from the ground state by absorption below $13.6 \: \rm{eV}$; 
the associated spectral lines belong to the Lyman and Werner band systems. 

\end{enumerate}

The fate of $\mHt$ in cold, neutral gas thus depends upon the flux of UV in 
the Lyman and Werner bands. Unfortunately, this 
is difficult to calculate accurately. Direct solution of the radiative 
transfer problem, even if decoupled from the hydrodynamics, is computationally
expensive. It is thus useful to explore analytic or semi-analytic alternatives
which can give reasonable estimates of the effects of the radiation.

One possible technique is that suggested by Omukai \& Nishi. They consider
a very simplified chemical model, in which the $\mHt$ abundance is set by the 
balance between formation via $\Hm$ and destruction by two-step 
photodissociation. The formation rate is limited by the rate at which $\Hm$ 
forms. This reaction has a rate coefficient \cite{dj}
\begin{equation}
\label{hmrate} 
k_{\rm form} = 1.0 \times 10^{-18} T \; \rm{cm}^{3} \: \rm{s}^{-1} 
\end{equation}
for gas at a temperature $T$. If destruction of $\Hm$ by other processes can 
be neglected, then this will also be the $\mHt$ formation rate. This 
approximation is justified as long as the fractional ionization is smaller
than $4 \times 10^{-3} T^{1/2}$ (J.\ Black, private communication), and
the density is greater than $0.045 (F_{\rm LW}/10^{-21})$, where $F_{\rm LW}$ 
is the average flux density in the Lyman-Werner bands, in units of $\rm{erg}
 \: \rm{s}^{-1} \: \rm{cm}^{-2} \: \rm{Hz}^{-1}$ \cite{mba}. The first of
 these conditions is certainly satisfied; the second becomes so only at a
distance of the order of $100 \: \rm{pc}$ or more from the nearest massive
star. However, as we see below, the size of the region affected by 
photodissociation in Omukai \& Nishi's model is greater still, so their 
approximation is justified.

The rate coefficient for $\mHt$ photodissociation can be written, in the 
optically thin limit, as \cite{db}
\begin{equation} 
k_{\rm 2step} = 1.13 \times 10^{8} F_{\rm LW} \; \rm{s}^{-1}. 
\end{equation}

With this simplified chemical model, $\mHt$ will reach chemical equilibrium 
on a timescale $t_{\rm dis} = k_{\rm 2step}^{-1}$. If this is significantly 
shorter than the timescales on which the flux, temperature, density or 
ionization 
vary then we can solve for the equilibrium fractional abundance of $\mHt$
\begin{eqnarray}
 x_{\mHt} & = & \frac{k_{\rm form}}{k_{\rm 2step}} x_{\rm e} n \nonumber \\
 & = & 8.8 \times 10^{-27} x_{\rm e} F_{\rm LW}^{-1} T n,
\end{eqnarray}
where $x_{\rm e}$ is the fractional ionization and $n$ is the total number 
density of the gas, in units of $\rm{cm}^{-3}$.
 
If self-shielding of the Lyman-Werner bands is unimportant then one can
 define the `region of influence' of a massive star to be the volume 
of gas surrounding it which satisfies two conditions; first, that the cooling 
time, given the equilibrium $\mHt$ abundance, is less than the free-fall 
timescale of the gas, and second, that this equilibrium abundance can be 
reached in less
than the lifetime of the central star.  Omukai \& Nishi consider the 
particular example of gas illuminated by a massive star with average 
luminosity density in the Lyman-Werner bands of 
$L_{LW} = 10^{24} \: \rm{erg} \: \rm{s}^{-1} \: \rm{Hz}^{-1}$. In this 
case, using their values of temperature, density and ionization, 
and assuming a uniform distribution of gas, the cooling condition implies a 
region of influence of radius
\begin{equation} 
r_{\rm cool} = 110 \; \rm{kpc},
\end{equation}
while the equilibrium condition implies a radius of 
\begin{equation}
r_{\rm eq} = 9.4 \; \rm{kpc}. 
\end{equation}
Both of these values are significantly larger
than the typical virial radius of a protogalactic halo,
\begin{equation} 
r_{\rm vir} = 200 \left(\frac{M}{10^{6} \: \rm{M}_{\odot}}\right)^{1/3} 
\: \rm{pc}, 
\end{equation}
(where we have followed Omukai \& Nishi  in adopting a gas number density
 $ n = 1 \: \rm{cm}^{-3}$). Hence, Omukai \& Nishi
 conclude that a single massive star can effectively suppress star formation
throughout the protogalaxy.

Self-shielding becomes important for gas at a radius $r$ from the star
when the intervening $\mHt$ column density, $N_{\mHt}$, 
exceeds $10^{14} \: \rm{cm}^{-2}$. In this case the situation is more 
complicated. The effect of self-shielding is to attenuate the Lyman-Werner 
flux. For the averaged flux this can be modelled by a shielding factor 
$f_{\rm sh}$, given with reasonable accuracy by \cite{db}
\begin{equation} 
f_{\rm sh} = \min \left[ 1, \left(\frac{N_{\mHt}}{10^{14}}
\right)^{-0.75}\right]. 
\end{equation}
However, unless the $\mHt$ abundance has already reached equilibrium, then the
 column density of $\mHt$ between the star and a given point, and hence the 
Lyman-Werner flux at that point, will change over time. Omukai \& Nishi
assume that the $\mHt$ abundance has reached equilibrium and use the 
 previously derived equilibrium abundance to calculate the distance from the 
star at which $N_{\mHt}=10^{14} \: \rm{cm}^{-2}$. Beyond this radius, 
$F_{\rm LW}$ drops sharply, and they consider this the effective 
boundary of the region of influence of the star. For the same example as 
 above, they find that
\begin{equation} 
r_{sh} = 0.97 \; \rm{kpc}.
\end{equation}
This is still comparable to the size of a typical protogalaxy. However,
 Omukai \& Nishi do not calculate the time taken to reach the equilibrium 
state. This may be significantly longer than in the optically thin case, as 
in the initially self-shielded gas $F_{\rm LW}$ and hence $k_{\rm 2step}$  
have values much lower than those in the equilibrium case. It is not clear that
this equilibrium can be reached within the lifetime of the star.

We attempt to remedy this difficulty by proposing a method for
estimating $t_{\rm dis}$ for optically thick gas, based upon a simple 
calculation of the time taken for the star to emit sufficient photons 
to dissociate the bulk of the $\mHt$. We outline our method in the next 
section.

\section{Estimating the dissociation timescale}
For a cloud containing a mass $M_{\mHt}$ of $\mHt$, we define the 
photodissociation timescale as
\begin{equation} 
t_{\rm{dis}} = \frac{M_{\mHt}}{\left|\dot{M}_{\mHt}\right|}, 
\end{equation}
where $\dot{M}_{\mHt}$ is the mass of $\mHt$ destroyed per second. In terms 
of the number of $\mHt$ molecules destroyed per second, $\dot{\cal N}_{\mHt}$ 
this is simply
\begin{equation} 
\dot{M}_{\mHt} = - m_{\mHt} \dot{\cal N}_{\mHt}, 
\end{equation}
where $m_{\mHt}$ is the mass of a hydrogen molecule.   
Now, $\dot{\cal N}_{\mHt}$ can be written as the product of four numbers
\begin{equation} 
\dot{\cal N}_{\mHt} = N_{\rm dis} f_{\rm inc} f_{\rm abs} f_{\rm dis}, 
\end{equation}
where $N_{\rm dis}$ is the number of photons capable of photodissociation 
produced per second, $f_{\rm inc}$ the fraction of these that are incident on 
the cloud,
$f_{\rm abs}$ the fraction of incident photons that are absorbed in the cloud, 
and $f_{\rm dis}$ the fraction of absorptions that lead to dissociation.
To calculate $t_{\rm dis}$ we thus need only calculate these four numbers. To
do this precisely is difficult, but fortunately we can estimate them 
relatively easily. 

\subsection{$N_{\rm dis}$}
Any photon with  sufficient energy can, in principle, cause the dissociation 
of an $\mHt$ molecule, although the probability of absorption for those 
photons with frequencies far from the centre of any Lyman-Werner line will be
small. To calculate $N_{\rm dis}$, therefore, we simply calculate the 
total number of photons between $11.15 \: \rm{eV}$ (the threshold for 
excitation to $B^{1}\Sigma^{+}_{u}$ from the ground state) and the Lyman 
limit at $13.6 \: \rm{eV}$. Higher energy photons will be blocked by 
neutral hydrogen, and are important only within the \hbox{H\,{\sc ii}} region.

\begin{figure}
\centering
\epsfig{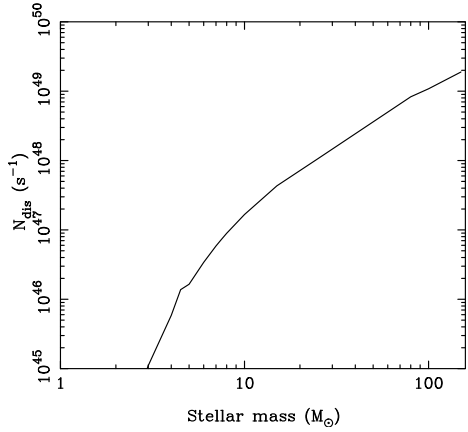}
\caption{The number of photons with energies between $11.15 \: \rm{eV}$ and
$13.6 \: \rm{eV}$ emitted per second by a metal-free star, as a function of 
stellar mass. The luminosities and effective temperatures of the stars are 
taken from Cojazzi et al.\ (2000).}
\label{ndis}
\end{figure}

For a single star, we have 
\begin{equation} 
N_{\rm dis}= \int^{\nu_{2}}_{\nu_{1}} \frac{L_{\nu}}{h \nu} \, d\nu, 
\end{equation}
where $L_{\nu}$ is the luminosity density of the star, 
$h\nu_{1}=11.15 \: \rm{eV}$ and $h\nu_{2}=13.6 \: \rm{eV}$. For primordial 
stars, the metallicity is effectively zero and we treat the stars as black 
bodies; for frequencies below the Lyman break, this should be a reasonable
approximation \cite{coj}. 
In this case 
\begin{equation}
L_{\nu} = 4 \pi R^{2}_{*} F_{\nu} 
\end{equation}
for a star of radius $R_{*}$,
where $F_{\nu} = \pi B_{\nu}$ is the flux at frequency $\nu$, $B_{\nu}$ 
is the Planck function
\begin{equation} B_{\nu} = \frac{2 h \nu^{3}}{c^{2}} 
\frac{1}{e^{h\nu/kT_{\rm{eff}}}-1},
\end{equation}
 and the stellar effective temperature is $T_{\rm{eff}}$.
We can write this in terms of the total luminosity $L_{*}$ as
\begin{equation} 
L_{\nu} = \frac{L_{*}}{\sigma T_{\rm{eff}}^{4}} \pi B_{\nu}. 
\end{equation}
$N_{\rm dis}$ is then given by
\begin{equation} 
N_{\rm dis} = \frac{L_{*}}{\sigma T_{\rm{eff}}^{4}} 
\int^{\nu_{2}}_{\nu_{1}} \frac{\pi B_{\nu}}{h\nu} \, d\nu.
\end{equation}
We obtain values of $L_{*}$ and $T_{\rm{eff}}$ as a function of stellar 
mass from Cojazzi et al.\ \shortcite{coj}, who fit an isochrone to a 
collection of ZAMS metal-free stellar models. It is then simple
to calculate $N_{\rm dis}$ as a function of stellar mass. We plot our results
 in figure~\ref{ndis}.

We do not incorporate the effects of stellar evolution into our model, as we 
expect them to be a relatively small correction. 

\subsection{$f_{\rm inc}$}
The fraction of the emitted flux incident on a 
region of interest depends upon the size of the region and its position with
respect to the star. For example, for a star in the centre of a spherically
symmetric cloud of gas, clearly $f_{\rm inc} = 1$. On the other hand, for a 
star illuminating a compact cloud from a distance, $f_{\rm inc} = \Omega/4\pi$,
where $\Omega$ is the solid angle subtended by the cloud, as seen from
the star.

\subsection{$f_{\rm abs}$}
The fraction of incident photons absorbed in the cloud is obviously frequency 
dependent; far more photons are absorbed at frequencies corresponding to the 
centres of the Lyman-Werner lines than in the wings. 
However, rather than tackle this frequency dependence directly, we instead
choose to approximate the effects of absorption in the set of Lyman-Werner 
lines by 
\begin{equation} 
f_{\rm abs} \simeq \frac{W_{\rm tot}}{W_{\rm max}}, 
\end{equation}
where $W_{\rm tot}$ is the total dimensionless equivalent width of the set of 
lines and  $W_{\rm max}=\rm{ln} \left(\frac{1110}{912} \right) \simeq 0.2$ 
is the dimensionless width of the range of wavelengths that contains all of the
Lyman-Werner lines.

The dimensionless equivalent width of a single line $i$ is 
\begin{equation} 
W_{i}=\int \frac{\rm{d}\nu}{\nu}(1-e^{-\tau_{\nu}}), 
\end{equation}
where $\tau_{\nu}$ is the optical depth at frequency $\nu$, and the integral 
is performed over the line profile. To calculate $W_{i}$, we assume that the 
line shape is well fit by a Voigt profile. We then need only a few
parameters to calculate the equivalent width. We take the oscillator 
strength of the transition and the frequency at line centre from Abgrall 
\& Roueff~\shortcite{ar} and
 the intrinsic broadening of the upper lines from Abgrall et al.\ 
\shortcite{abp}. We also need 
to know the Doppler broadening parameter,
\begin{equation} 
b=\sqrt{\frac{2kT}{m_{\mHt}}+ v_{\rm t}^{2}},
\end{equation}
where $v_{\rm t}$ is the microturbulent velocity.
Clearly, we will not in general know the appropriate value of $b$ to use; 
however, it is possible to place some limits on it.
Our lower limit, $b_{\rm min}=1.3 \: \rm{km} \: \rm{s}^{-1}$, corresponds to 
pure 
thermal broadening at a temperature of $200 \: \rm{K}$. This is approximately 
the lowest 
temperature at which $\mHt$ cooling is effective; the amount of gas cooler than
this is expected to be negligible, unless HD cooling should prove to be 
important.  
We obtain an approximate upper limit by
assuming thermal broadening at $6000 \: \rm{K}$ (a higher temperature implies 
significant collisional dissociation), plus a similarly-sized contribution 
from microturbulence. This gives 
$b_{\rm max} \simeq 10 \: \rm{km} \:\rm{s}^{-1}$, and is a strong upper limit. 
Given some value of $b$ within this range, it is then straightforward 
to  calculate $W_{i}$ as a function of $N_{\mHt}$ by means of the accurate 
numerical approximations given in Rodgers \& Williams~\shortcite{rw}. 

For a set of lines, one might naively expect that
\begin{equation} 
W_{\rm tot} = \sum_{i} W_{i}, 
\end{equation}
ie.\ that the total dimensionless equivalent width is simply the sum of the 
widths
of the individual lines. Unfortunately, this expression behaves unphysically 
in the high column density limit, as it does not account for the effects of 
line overlap. Clearly $W_{\rm tot} \leq W_{\rm max}$ -- the cloud cannot 
absorb more light than is actually there. To correct for the effects of overlap
we follow the procedure suggested in Draine \& Bertoldi~\shortcite{db}, using 
a modified equivalent width defined 
as\footnote{Note that while Draine \& Bertoldi also included the effects of 
absorption in
the hydrogen Lyman series in their calculation of the effects of overlap, it 
is clear from their figure 6 that these generally have a very small effect on
$\tilde{W}_{\rm tot}$.}
\begin{equation} 
\tilde{W}_{\rm tot} = W_{\rm max} [1- \exp (-W_{\rm tot}/W_{\rm max})]. 
\end{equation}
  It is clear from this that $\tilde{W}_{\rm tot} \rightarrow W_{\rm max}$ as 
$W_{\rm tot} \rightarrow \infty$. We then have
\begin{equation} 
f_{\rm abs}  =  \frac{\tilde{W}_{\rm tot}}{W_{\rm max}} = 1- \exp 
(-W_{\rm tot}/W_{\rm max}). 
\end{equation}

\begin{figure}
 \centering
\epsfig{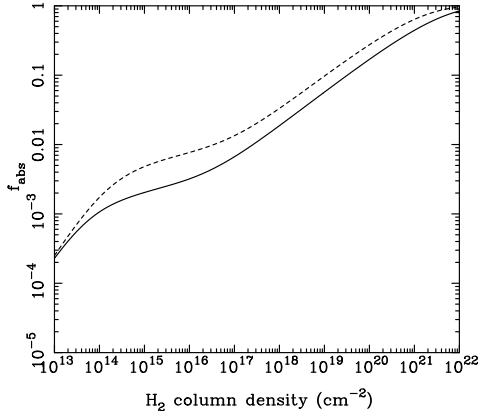}
\caption{The fraction of radiation in the Lyman-Werner bands that is absorbed
by the gas, for Doppler parameter $b=b_{\rm min}$, for an ortho- to para-
ratio of zero (solid line) or three (dashed line). }
\label{width}
\end{figure}

\begin{figure}
\centering
\epsfig{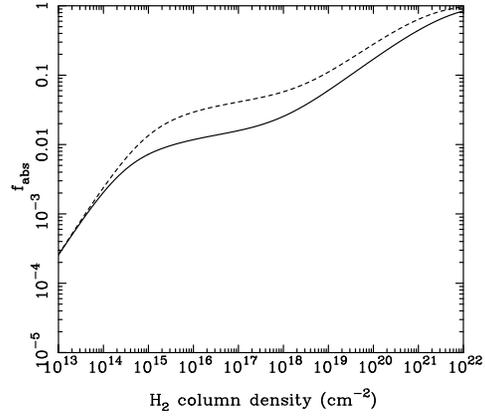}
\caption{The same as figure 2, but with $b=b_{\rm max}$}
\label{width2}
\end{figure}

Finally, to calculate $\tilde{W}_{\rm tot}$ correctly, we must know 
which lines to include. As previously noted, we expect almost all of the $\mHt$
to be in its ground state. The ortho- to para-hydrogen ratio is 
not known \emph{a priori}, but there is good reason
to believe that para-hydrogen will predominate \cite{aanz}.
Assuming that there is \emph{no} ortho-hydrogen, 
it is easy to calculate $\tilde{W}_{\rm tot}$, and hence $f_{\rm abs}$,
as a function of $N_{\mHt}$. We plot this in figure~\ref{width} for  
$b=b_{\rm min}$ and in figure~\ref{width2} for $b=b_{\rm max}$.
For comparison, in both of these figures we also plot $f_{\rm abs}$ 
obtained with the assumption of an ortho- to para- ratio of three to one. 

We see that uncertainties in the ortho- to para- ratio will affect our 
calculated value of $f_{\rm abs}$ for all but the largest or smallest column 
densities.
However, the error introduced by this is relatively small -- the difference
between our two extreme cases is never more than a factor of two. The 
uncertainty in $b$ is potentially more important, but its effects are 
restricted to the square-root portion of the curve of growth, between 
$N_{\mHt}=10^{14} \: \rm{cm}^{-2}$ and $N_{\mHt}=10^{18} \: \rm{cm}^{-2}$.
Even in this region, the difference is only a factor of five
 and outside of this region it quickly becomes unimportant.

\subsection{$f_{\rm dis}$}
The fraction of excitations that are followed by dissociation depends upon the
 vibrational state of the excited level; values for individual levels are 
given in Abgrall et al.\ \shortcite{abp}. Consequently, $f_{\rm dis}$ will 
depend upon the shape 
of the incident spectrum, and hence the column density of $\mHt$. Fortunately,
this dependence is not large. Draine \& Bertoldi show that, for a power law 
spectrum,
the dissociation probability varies between 0.1 and 0.2 with a mean value of 
0.15. Moreover, changing the shape of the spectrum changes this value by only
a few percent.

The situation is complicated, however, by the photons emitted by those $\mHt$
molecules which do not photodissociate. The majority of decays are initially 
to highly excited vibrational states, and the consequent photons have energies
below $11.15 \: \rm{eV}$ and play no further part in photodissociation.
A fraction, however, decay to low-lying vibrational states and
produce photons energetic enough to cause photodissociation; 
indeed, some will decay directly back to the ground state, producing photons
coincident with ground state absorption lines.

The net effect is to scatter some fraction of the incident radiation, while at
the same time redistributing it in frequency space. To investigate the 
importance of this effect, we first assume that the dissociation 
probability after each absorption remains 0.15; this need not be the case, but
is a reasonable first approximation. Next, we divide the emitted photons into
three classes. First, we have those with energies less than $11.15 \: \rm{eV}$;
these can simply be ignored. Second, we have those produced by decay to an 
excited state, but with energies greater than $11.15 \: \rm{eV}$. These will
not coincide with ground-state absorption lines, and thus will only be 
re-absorbed in the limit of high column densities, $N_{\mHt} \gg 10^{21} \:
\rm{cm}^{-2}$, when line overlap becomes important. We denote the fraction of 
decays producing photons of this type by $p_{\rm LW}$. This will depend upon 
the shape of the incident spectrum, but as in the case of $f_{\rm dis}$ we
expect this dependence to be small. A simple unweighted average over all of the
excited states available from the ground state gives $p_{\rm LW} = 0.16$, which
we take as a reasonable estimate. 
Finally, some fraction of the photons are produced by direct decay to
 the ground state. Following Draine \& Bertoldi we denote this fraction by 
$p_{\rm ret}$. This is sensitive to the ortho- to para-hydrogen ratio, but
 assuming the standard equilibrium value of three-to-one and performing
an unweighted average over accessible states gives us $p_{\rm ret}=0.08$.
Alternatively, if we have pure para-hydrogen, $p_{\rm ret}=0.04$. We adopt the
 former value hereafter.

Armed with these values, we can now calculate the size of the correction we
must make to $f_{\rm dis}$. The largest effect is found in the limit of high 
column densities. For $N_{\mHt} \gg 10^{21} \:\rm{cm}^{-2}$, line overlap
implies that all photons within the Lyman-Werner bands will be absorbed. 
In this case, we can write $f_{\rm dis}$ as:
\begin{equation}
 f_{\rm dis} = 0.15 \left(\frac{1}{1-p_{\rm LW}-p_{\rm ret}}\right) 
\end{equation}
With our values of $p_{\rm LW}$ and $p_{\rm ret}$ this implies that 
$f_{\rm dis} \simeq 0.2$. Thus the effect will never be large; indeed, since
the size of the correction is comparable to our uncertainty in $f_{\rm dis}$
we are justified in ignoring this effect in the calculations that follow.
 
\subsection{Calculating the dissociation timescale}
Putting together our various estimates, we have:
\begin{eqnarray}
 t_{\rm{dis}} & = & \frac{M_{\mHt}}{\left|\dot{M}_{\mHt}\right|} \nonumber \\
 & = & \frac{x_{\mHt} M}{m_{\mHt} \dot{\cal N}_{\mHt}},
\end{eqnarray}
where
\begin{equation}
\dot{\cal N}_{\mHt} =  N_{\rm dis} f_{\rm inc} f_{\rm abs} f_{\rm dis} 
\end{equation}
Here $x_{\mHt}$ is the fractional abundance of $\mHt$ and $M$ is the total 
mass of the cloud.  
Combining the two equations, and writing $M$ in units of $M_{\odot}$, we
have
\begin{equation} 
t_{\rm dis} \simeq 2 \times 10^{49} \frac{x_{\mHt} M}{N_{\rm dis} 
f_{\rm inc} f_{\rm abs} f_{\rm dis}} \: \rm{yr}. 
\end{equation}
For the example of a $25 M_{\odot}$ star, $N_{\rm dis} 
\simeq 10^{48}$ and 
\begin{equation} 
t_{\rm dis} \simeq 20 \frac{ x_{\mHt} M}{f_{\rm inc} f_{\rm abs}
 f_{\rm dis}} \left( \frac{N_{\rm dis}}{10^{48}} \right)^{-1} \: \rm{yr}. 
\end{equation}

\section{Applications}
As an example, we apply our method to a very simple protogalactic
model, consisting of two components:
a spherically symmetric, diffuse
component (the halo), in which are embedded clumps of a significantly denser
component. Although recent numerical simulations offer some 
support for this 
picture, doubts inevitably remain concerning the general 
applicability of the results. Moreover, observations of the local molecular 
clouds suggest that it may be an oversimplification. Indeed, there is 
evidence that molecular clouds have a fractal structure \cite{elm}, although
this remains controversial \cite{bw}. Nevertheless, our model has the 
advantage of simplicity, and should be a reasonable guide.

\subsection{The spherical halo}
We first consider the simple case of a single star at the centre of a spherical
halo. As a representative example, we consider a $25 M_{\odot}$ star, which 
gives $N_{\rm dis}=10^{48} \: \rm{s}^{-1} $, although it is straightforward 
to rescale our results for a different stellar mass. Now, spherical 
symmetry implies $f_{\rm inc} = 1$, so it remains
only to specify the mass, fractional $\mHt$ abundance and 
$\mHt$ column density of the
halo. From these we can then calculate $f_{\rm abs}$ and $f_{\rm dis}$.

We can place a number of constraints upon these parameters.
 The fractional abundance is the most straightforward. A lower limit upon its 
value comes from the requirement that there be enough $\mHt$ to cool the gas 
effectively. The required fractional abundance has been estimated by  
Tegmark et al.\ \shortcite{teg} to be approximately $ 5 \times 10^{-4}$. 
We err on the side of caution and adopt a lower limit of 
$10^{-4}$. An upper limit can be obtained via a method suggested by 
Nishi \& Susa~\shortcite{ns}. They note that the maximum abundance of $\mHt$ 
that can form in 
 a halo is approximately the amount that forms within some
critical timescale. This critical timescale is temperature dependent, and is
the shortest of the recombination, cooling and dissociation timescales. 
For low ionization gas, this typically gives a maximum abundance of order 
$10^{-3}$.\footnote{This does not apply to very dense gas, where
$\mHt$ forms predominantly via three-body processes and the $\mHt$ fractional
 abundance is typically near unity.}

We can also constrain the halo mass by a cooling argument, given the
redshift at which it forms. This depends upon the cosmological model.
A frequently adopted value is $z=30$, appropriate to the case of a rare 
halo in an SCDM
cosmology. However, values in the range $10<z<100$
are plausible (depending upon the rarity of the halo at formation) if we do
not limit ourselves to a particular cosmological model.

Given the redshift, a lower limit upon the halo mass $M$ comes from the 
requirement that the 
virial temperature of the halo be high enough to form sufficient $\mHt$ for
effective cooling. This has also been studied by Tegmark et al.
They find that the
critical mass is redshift dependent, increasing at low redshifts. In
the range that we consider, their minimum value is $M_{\rm min}=10^{5} 
{\rm M}_{\odot}$. Adopting a gas to dark matter ratio $f_{\rm b}=0.1$, this 
corresponds to a minimum gas mass of $10^{4} \: {\rm M}_{\odot}$.

An upper limit on the mass comes from the fact that halos with 
virial temperatures greater than $10^{4} \: \rm{K}$ will cool effectively via  
Lyman-$\alpha$ emission. Although it is unclear what role $\mHt$ cooling
plays in such halos, and whether it is necessary for star formation, it is
often assumed that Lyman-$\alpha$ cooling suffices. As a detailed examination 
of this issue lies beyond the scope of this paper, we also make this 
assumption, and do not consider larger halos.
The mass corresponding to a virial temperature of $10^{4}$K is given by
\begin{equation} 
M_{\rm max} = 2.2 \times 10^{9} h^{-1} (1+z)^{-3/2} \: {\rm M}_{\odot}, 
\end{equation}
where $100h \: \rm{km} \: \rm{s}^{-1} \: \rm{Mpc}^{-1}$ is the Hubble constant.
In the range of redshifts we consider, this corresponds to a maximum mass of
approximately $10^{8} \: {\rm M}_{\odot}$ and hence a maximum gas mass of 
$10^{7} \: {\rm M}_{\odot}$.

Given the mass and $\mHt$ abundance, we next specify the $\mHt$ column density,
$N_{\mHt}$, by choosing a halo density profile. The simplest choice is a 
uniform sphere. In this case the spherical collapse model suggests that the 
resulting virialized halo will have an 
overdensity of $ \delta = 18\pi^{2}$ with respect to the cosmological 
background at the 
time of collapse. This is equivalent to a total number density of
\begin{equation} 
n_{\rm vir} = 2.1 \times 10^{-5} \left(\frac{\Omega_{\rm b} h^{2}}
{0.0125}\right)  (1+z)^{3} \; \rm{cm}^{-3}, 
\end{equation}
for a neutral gas of primordial composition, and hence an $\mHt$ number 
density 
\begin{equation}
n_{\mHt} = x_{\mHt} n_{\rm vir}.
\end{equation}
Since we also know the mass, it is now trivial to solve for the virial radius
of the halo, $r_{\rm vir}$, and hence for $N_{\mHt}$. We find that, for
a star at the centre of the halo, the $\mHt$ column density along a radial ray 
is
\begin{equation}
N_{\mHt}= 4.7 \times 10^{15} x_{\mHt} M^{1/3} \left(\frac{\Omega_{\rm b} 
h^{2}}{0.0125}\right)^{2/3} (1+z)^{2} \; \rm{cm}^{-2}. 
\end{equation}

Unfortunately, the uniform sphere is an unrealistic model; both theory
and computer simulation suggest that real protogalactic clouds will be 
centrally concentrated. To address this, we consider a second model, that of a
truncated isothermal sphere \cite{shap}. 
This has a core overdensity $\delta=1.796
 \times 10^{4}$, a core radius $r_{0}=0.038r_{\rm vir}$ and a truncation 
radius $r_{\rm t}=1.108r_{\rm vir}$ (corresponding to $r_{\rm t}=29.4r_{0}$).
Over much of its radius, the profile  is approximately isothermal. Comparing 
the two column densities, we find that
\begin{equation} 
N_{\rm T} = 11.3 N_{\rm U}, 
\end{equation}
where $N_{\rm T}$ and $N_{\rm U}$ are the column densities for the truncated 
isothermal sphere and uniform sphere respectively.
The uniform sphere likely underestimates the column density by an 
order of magnitude.
However, in the truncated isothermal sphere model, a
significant fraction of this column density comes from the small, dense core.
Outside of the core, the uniform sphere model may give a  better estimate.
We therefore continue to consider both models, with the expectation that the
uniform sphere model will provide us with a strong lower limit on 
$t_{\rm dis}$, while the truncated isothermal model should provide a 
reasonable upper limit.

There remain two parameters to specify, related to the microphysics of the gas.
These are the Doppler parameter $b$ and the ortho- to para-hydrogen ratio. 
Both of these must be regarded as rather uncertain. We consider two sets of 
extreme values. For the uniform sphere model, we assume $b=b_{\rm min}$ and 
an ortho- to para- ratio of zero (ie.\ no ortho-hydrogen),
while for the truncated isothermal sphere model, we assume $b=b_{\rm max}$ 
and an ortho- to para- ratio of three.
Our two models now give strong upper and lower limits on $t_{\rm dis}$.

Applying our calculations to the concrete example of a halo of gas mass 
 $M=10^{5} \: {\rm M}_{\odot}$, with an $\mHt$ abundance of $5 \times 10^{-4}$
that formed at $z=30$, we find that
\begin{equation} 
8.3 \times 10^{4} \: \rm{yr} < t_{\rm dis} < 9.5 \times 10^{5} \: \rm{yr}. 
\end{equation}
To put these figures into context, the massive stars we are considering will
have main sequence lifetimes that are typically of the order of a few Myr, so
in this particular case cooling will cease before the end of the star's life. 

\begin{figure}
\centering
\epsfig{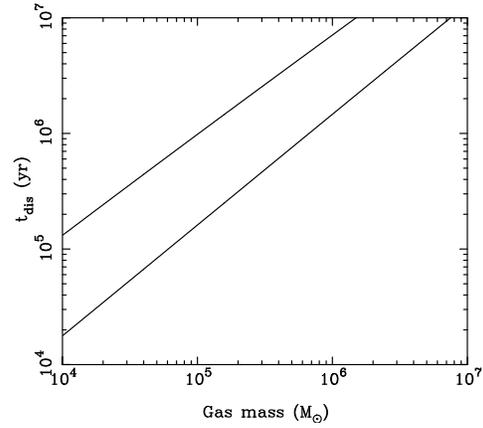}
\caption{Upper and lower limits on the photodissociation timescale, plotted as
a function of the mass of gas in the halo, for halos forming at $z=30$ with 
$\mHt$ abundance $x_{\mHt} = 5 \times 10^{-4}$.}
\label{emm}
\end{figure}

\begin{figure}
\centering
\epsfig{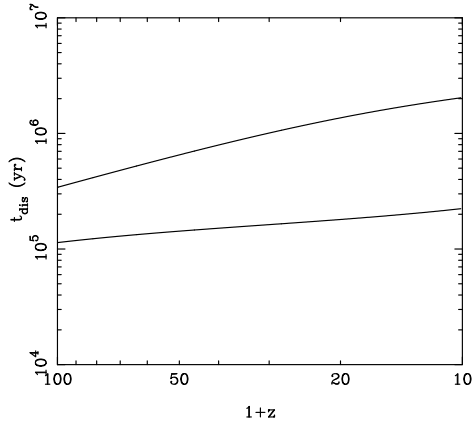}
\caption{As figure~\ref{emm}, but examining the effects of varying the 
redshift of formation. The gas mass is fixed at $10^{5} M_{\odot}$, and the 
$\mHt$ abundance is again $x_{\mHt} = 5 \times 10^{-4}$.}
\label{zed}
\end{figure}

\begin{figure}
\centering
\epsfig{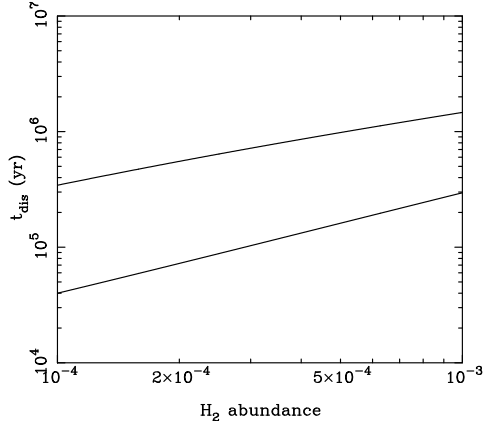}
\caption{As figure~\ref{emm}, but examining the dependence upon $\mHt$ 
abundance. The gas mass is $10^{5} M_{\odot}$ and the halos form at a redshift
$z=30$.}
\label{ft}
\end{figure}

We next proceed to vary the parameters individually, to examine their relative
importance. We take as a fiducial model the example above, 
 vary $M$, $x_{\mHt}$ and $z$ over their allowed range, 
and plot the results in figures~\ref{emm} to \ref{ft} 
respectively. 
Several points are immediately apparent.

Firstly, the most important parameter for determining $t_{\rm dis}$ is the 
mass of
gas in the halo. Its importance is due to the weak dependence of $N_{\mHt}$
upon mass. As the mass of the halo increases, the mass of $\mHt$ within it 
increases proportionately, but the column density increases at a much slower 
rate. This means that a large increase in mass leads to only a small increase
in the amount of radiation absorbed, and hence the increase in $t_{\rm dis}$ 
is large. Moreover, our constraint upon the halo mass is much poorer than those
upon redshift or $\mHt$ abundance.

Another conclusion that can be drawn is that changes in $z$ (which are 
equivalent to changing the $\mHt$ column density while keeping
 the mass of $\mHt$ in the halo fixed) have only a small effect. 
This is because the $\mHt$ column densities for the halos considered 
generally correspond to
equivalent widths in the square-root portion of the curve of growth. Thus, 
large changes in $N_{\mHt}$ lead to only small changes in $f_{\rm abs}$ and 
$t_{\rm dis}$.

It is clear from figure~\ref{emm} that in the smallest halos photodissociation
 equilibrium will be reached quickly, within a small fraction of the stellar 
lifetime. We can therefore apply the results of Omukai \& Nishi 
with confidence to such halos.
 As the mass of gas in the halo increases, however, $t_{\rm dis}$ becomes 
steadily more important, and the assumption of equilibrium breaks down. For 
example, when the gas mass is $10^{6} M_{\odot}$, the \emph{lower} limit on 
$t_{\rm dis}$ is approximately $10^{6}$ years -- a significant fraction of the 
stellar lifetime. We cannot safely apply Omukai \& Nishi's results to such 
halos, 
and they may undergo significant cooling right up until the point at which the
massive star becomes a supernova. 

This situation changes somewhat if we increase $N_{\rm dis}$, as would occur if
a more massive star formed, or if several massive stars formed at 
approximately the same time. In 
this case our estimated values of $t_{\rm dis}$ would be correspondingly 
lower, and cooling would be suppressed much sooner. 
Nevertheless, even in this case we cannot conclude that UV feedback
necessarily implies a low star formation efficiency, as much gas may already
have cooled and collapsed into dense clumps by the time the first star forms.
To properly ascertain the effects of the ultraviolet radiation
on star formation efficiency, we must also consider the fate of $\mHt$ 
within these dense clumps.

\subsection{Dense clumps} 
The second component of our protogalactic model consists of
dense clumps of gas embedded in the diffuse halo.
 The free-fall timescale of such clumps, $t_{\rm ff}$, will be considerably 
shorter than that of the halo. In the absence of significant 
pressure support, we would expect the clumps to collapse quickly and presumably
to form stars on a similar timescale. This will only be true so long as the 
clumps remain able to cool efficiently. We assume that
clumps will collapse and form stars only if they retain their $\mHt$ until 
collapse, ie.\
only if $t_{\rm dis} > t_{\rm ff}$.

To see whether clumps are capable of preserving their $\mHt$ for long enough to
form stars, we need to calculate both of these
timescales. The free-fall timescale is straightforward; for gas of density
$\rho$ it is simply
\begin{eqnarray} 
t_{\rm ff} & = & (G\rho)^{-1/2}  \nonumber \\
 & \simeq & 10^{8} n^{-1/2} \: \rm{yr}. 
\end{eqnarray}
Comparing this with a typical massive star lifetime of a few Myr, we see 
 that we need not consider clumps with densities lower than $10^{3} \:
\rm{cm}^{-3}$;  at such densities, clumps will not have collapsed by the 
time
the star becomes a supernova. Their subsequent evolution will 
be shaped by their interaction with the supernova remnant, a topic which lies
beyond the scope of this paper.

Our calculation of $t_{\rm dis}$ for denser clumps is complicated by the fact 
that at high densities our assumption that all $\mHt$ molecules will be in the
 ground state is no longer valid. At densities above some critical value 
$n_{\rm cr}$, collisional excitation and de-excitation dominate, and the 
populations of excited states approach their LTE values. The value of the 
critical density depends upon the particular excited state, as well as the 
temperature, but typically $n_{\rm cr} \sim 10^{4} \: \rm{cm}^{3}$ for the 
lower rotational states \cite{bpf}. Vibrational states have a much higher
critical density ($n_{\rm cr} \sim 10^{8} \: \rm{cm}^{-3}$), and in any case
will have a negligible population at $1000 \: \rm{K}$.

In calculating $t_{\rm dis}$ for dense clumps we therefore assume that the 
rotational states are in local thermodynamic equilibrium, with a temperature 
of $1000 \: \rm{K}$. The main effect of this is to increase $f_{\rm abs}$ (and
hence decrease $t_{\rm dis}$), typically by a factor of two to 
three.

Next, we note that for dense clumps $f_{\rm inc} \ll 1$; individual clumps
 receive only a small fraction of the total flux from the star. Thus, despite
having a larger value of $f_{\rm abs}$ and a much smaller mass of $\mHt$ than 
the typical halo, it does not necessarily
follow that a clump will have a shorter dissociation timescale.

For a clump of angular size $\Omega$, as seen from the star, we have 
$f_{\rm inc}=\Omega/4\pi$. In the particular case of a spherical clump, in the
 small angle approximation, 
\begin{equation} 
f_{\rm inc}= \frac{R^{2}}{4 D^{2}}, 
\end{equation}
where $R$ is the radius of the clump, and $D$ is the distance from clump to 
star.
 For simplicity we do not consider ellipsoidal or filamentary clumps; however,
 it would be straightforward  to extend our analysis to them.

Modelling the density structure of the clumps is problematic; we have little 
theoretical guidance. Consequently, we adopt the somewhat unrealistic model 
of a uniform density sphere. This will minimize $t_{\rm dis}$ and maximize 
$t_{\rm ff}$ for a given clump. We can then be confident that if our 
model clumps collapse, so would more realistic clumps.

For a given clump we specify four parameters: its mass 
(in units of ${\rm M}_{\odot}$), density, $\mHt$
abundance and  distance from the star. From these quantities, we 
calculate the clump radius
\begin{equation} 
R = 6.2 \times 10^{18} M^{1/3} n^{-1/3} \; \rm{cm}, 
\end{equation}
 and the $\mHt$ column density along a radial ray passing through the centre 
of the clump,
\begin{equation} 
N_{\mHt} = 1.2 \times 10^{19} x_{\mHt} M^{1/3} n^{2/3} \; \rm{cm}^{-2}. 
\end{equation}
From these we calculate $f_{\rm inc}$, $f_{\rm abs}$ and $f_{\rm dis}$ and 
thence $t_{\rm dis}$. We find that
\begin{equation} 
t_{\rm dis} = 20 x_{\mHt} M^{1/3} n^{2/3} D^{2} \left(\frac{N_{\rm dis}}
{10^{48}} \right)^{-1} f_{\rm abs}^{-1} f_{\rm dis}^{-1} \; \rm{yr}, 
\end{equation}
where $D$ is in parsecs.

We immediately see that the dependance of $t_{\rm dis}$ upon clump mass is 
weak,
unlike the case of the halo. This is due to the scaling of $f_{\rm inc}$ with
 clump mass; since massive clumps are larger, they absorb more of the 
flux, which, to a great extent, offsets the fact that they contain more $\mHt$.
This weak dependence is rather fortunate, as we know very little about the 
mass spectrum of clumps to be found within protogalaxies.

The dependence of $t_{\rm dis}$ upon $x_{\mHt}$ is stronger, but we do not 
expect the $\mHt$ abundance in clumps to differ much from that in the halo; 
consequently, the same tight constraints apply, and the overall 
uncertainty in $t_{\rm dis}$ is small. In very dense clumps, as noted 
previously, this does not apply -- three-body formation of $\mHt$ allows much 
higher abundances to form. However, this process only becomes important above 
a number density of $10^{8} \: \rm{cm}^{-3}$; as we show below, in clumps of 
this density, $\mHt$ will in any case survive until collapse.

\begin{figure}
\centering
\epsfig{figure=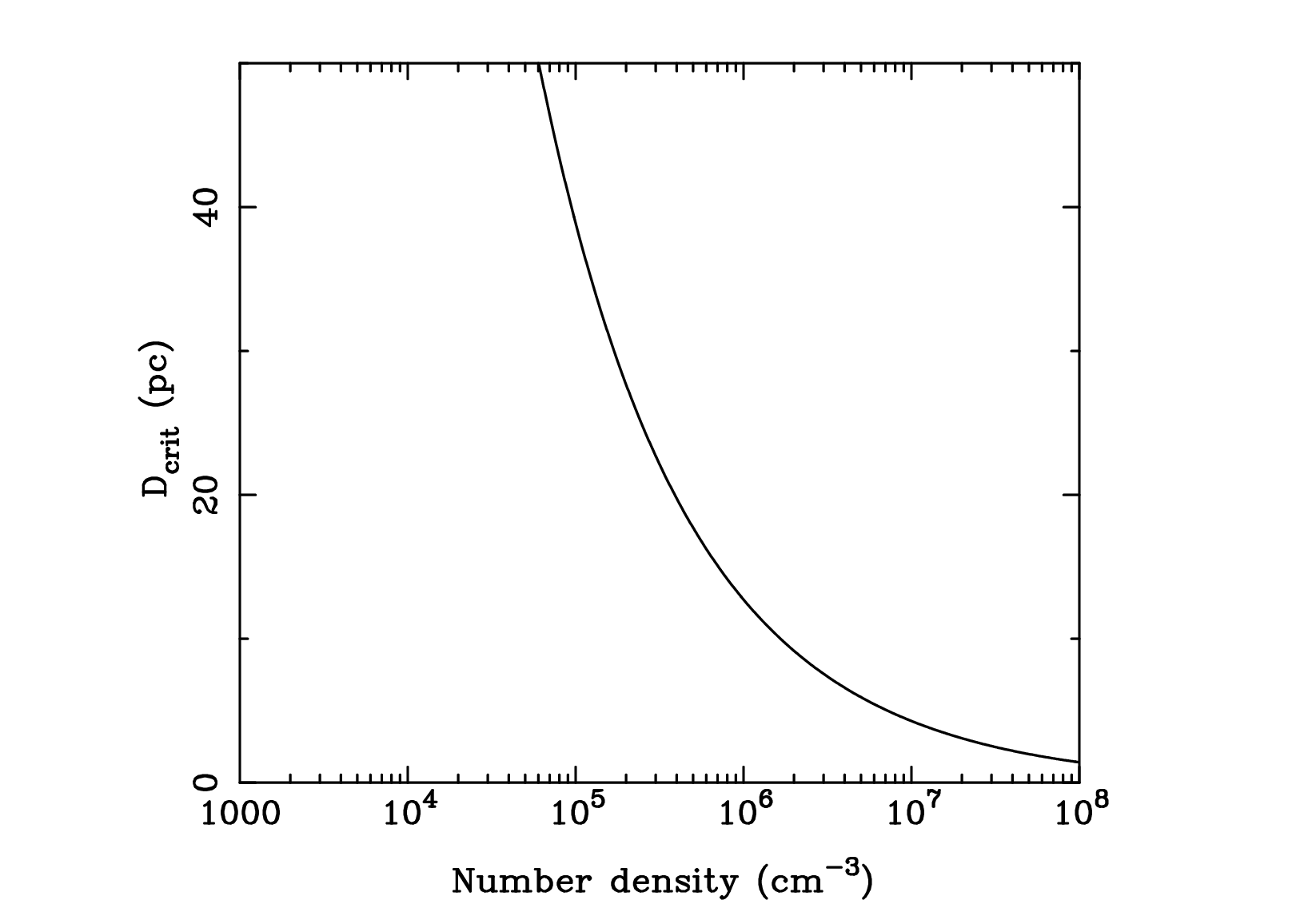,width=5.8cm,angle=270,clip=}
\caption{The distance at which the dissociation and free-fall timescales are 
equal, plotted as a function of clump density, for a clump with $x_{\mHt} 
M^{1/3} = 10^{-4}$.}
\label{small}

\centering
\epsfig{figure=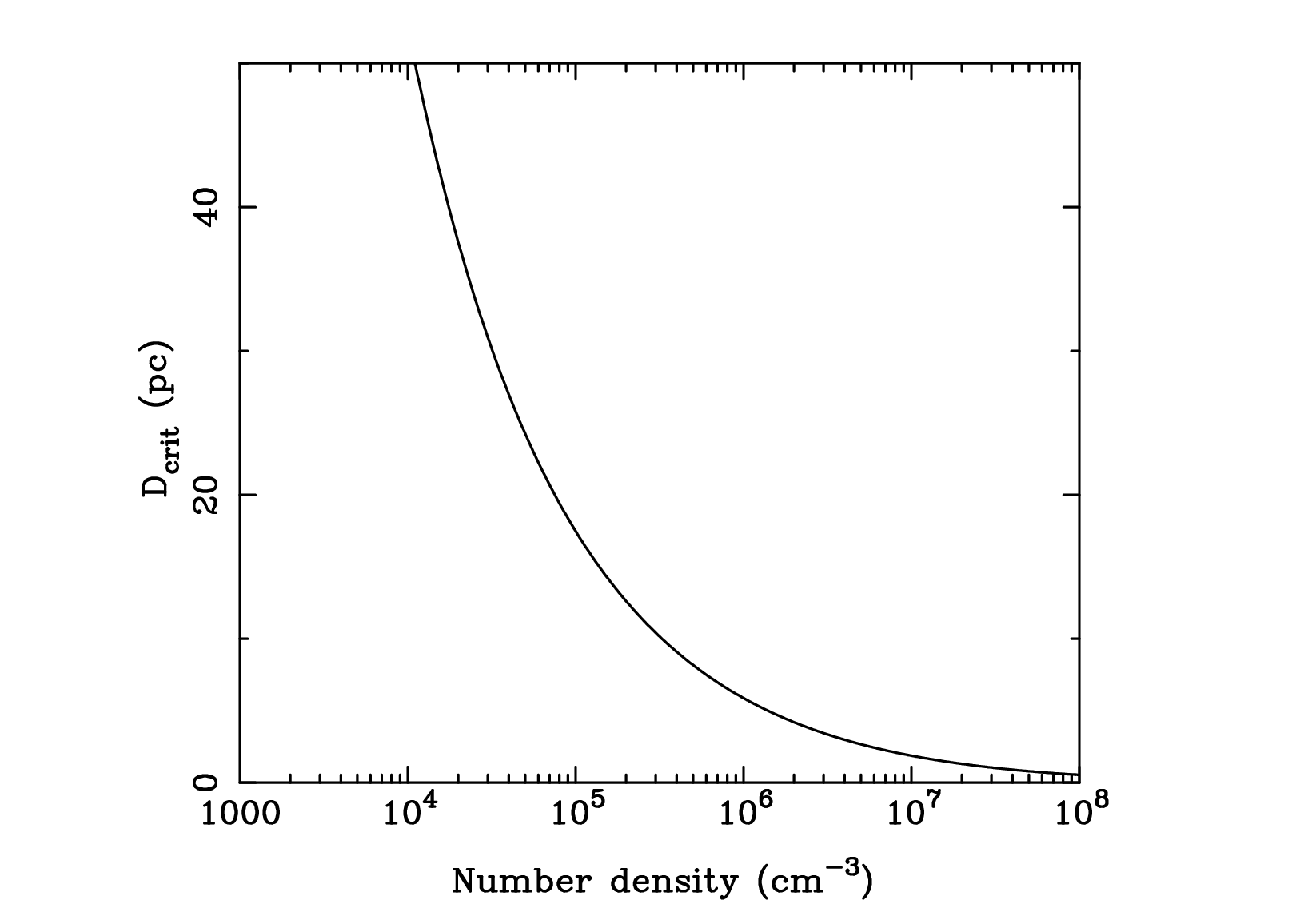,width=5.8cm,angle=270,clip=}
\caption{As figure~\ref{small}, but for a clump with $x_{\mHt} M^{1/3} = 
10^{-3}$.}
\label{medium}

\centering
\epsfig{figure=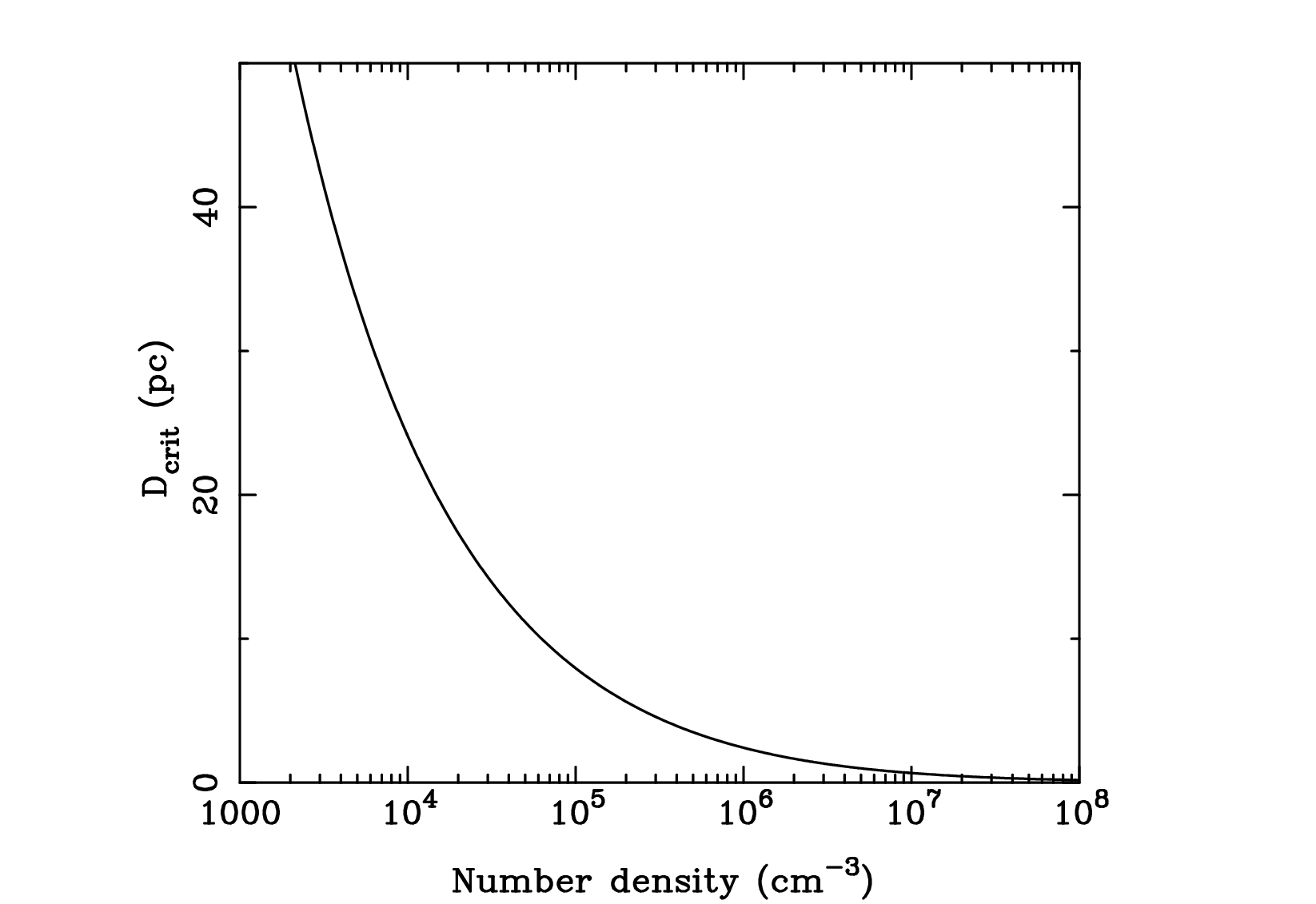,width=5.8cm,angle=270,clip=}
\caption{As figure~\ref{small}, but for a clump with $x_{\mHt} M^{1/3} = 
10^{-2}$.}
\label{large}
\end{figure}

The most important parameters for determining the fate of a clump are thus its
density and its distance from the star. For a given mass and $\mHt$ abundance,
 we can set $t_{\rm dis}$ equal to $t_{\rm ff}$ and solve for the critical 
distance $D_{\rm crit}$ at which $\mHt$ survives until collapse; this will be 
a function of clump density.
We examine three cases, with different values of $M$ and $x_{\mHt}$.
Rather than varying these quantities separately, we instead vary the 
combination
$x_{\mHt} M^{1/3}$, as it is in this form that the mass and $\mHt$ abundance 
appear in our expressions for the 
$\mHt$ column density and for $t_{\rm dis}$.  
We examine cases corresponding to values of $10^{-4}$, $10^{-3}$ and $10^{-2}$
for $x_{\mHt} M^{1/3}$; we plot  
the results in figures~\ref{small}, \ref{medium} and \ref{large} 
respectively. These values
 cover the range of clump masses and $\mHt$ abundances that we 
would reasonably expect to encounter.

The qualitative behaviour is clear. High density clumps easily preserve 
their $\mHt$ until collapse, unless they happen to be very close to the 
star. Lower density clumps, on the other hand, are prevented from collapsing
at significant distances from the star -- to put these values for 
$D_{\rm crit}$ into context, the virial radius of a 
$10^{6} \, {\rm M}_{\odot}$ halo collapsing at $z=30$ is about 
$100 \: \rm{pc}$. 

To quantify the effects of ultraviolet feedback upon star formation within 
clumps is rather harder, as in this case we are not interested in the absolute
 value of $D_{\rm crit}$ so much as its size relative to the size of the 
star-forming region. Nevertheless, a rough guide to the expected effects would
come from assuming that stars will form only in clumps denser than some 
threshold, say $n = 10^{6} \: \rm{cm}^{-3}$, for which 
$D_{\rm crit}$ is small compared to the size of the protogalaxy.    
If we increase $N_{\rm dis}$ (by forming more massive stars, for example), 
then this density threshold will increase. On the other hand, we expect our 
model to give us an upper limit on $D_{\rm crit}$ and hence on 
the density threshold; realistic clumps may well survive much nearer to 
massive stars than our model clumps do.

\section{Complications}
In order to construct a simple approximation, we have omitted a number of
 complicating features. Some of these omissions call for further comment.
 These are the heating of the gas during photodissociation, 
the effects of circumstellar material, the 
contribution to $\mHt$ dissociation made by \hbox{H\,{\sc ii}} regions and 
stellar winds and the possible effects of dust.

\subsection{Gas heating due to photodissociation}
Rather than simply suppressing cooling, $\mHt$ photodissociation will in 
actual fact lead to heating of the gas, as some of the energy of the 
Lyman-Werner photon is transferred to the kinetic energy of the hydrogen atoms.
However, it is easy to show that the the effects of this heating upon our 
protogalactic model will be small. A single photodissociation typically 
produces $0.4 \: \rm{eV}$ of heat \cite{bd}, and thus destruction of all of 
the $\mHt$ within a cloud will increase the temperature (in the absence of 
any other heating or cooling) by
\begin{eqnarray}
\Delta T & = & \frac{6.4 \times 10^{-12}}{k}x_{\mHt} \: \rm{K} \nonumber \\
 & \simeq & 50000 x_{\mHt} \: \rm{K}.
\end{eqnarray}
As our upper limit on $x_{\mHt}$ is $10^{-3}$, this corresponds to a maximum 
temperature increase of roughly $50 \: \rm{K}$. This is too small to 
significantly affect our results, and thus we are justified in ignoring this 
effect.

\subsection{Circumstellar material}
Since we expect stars to form within dense clumps of gas, it is reasonable to 
assume that they will still be surrounded by large 
amounts of molecular gas after they have formed. If this is so, then this gas 
may absorb most of the 
Lyman-Werner photons emitted by the star, in which case dissociation in the 
halo or in other dense clumps will be reduced until the bulk of the 
circumstellar $\mHt$ is destroyed.

If this material is distributed in a spherically symmetric fashion around the 
newly formed star, then it is quite easy to estimate its effects. There are 
two possibilities. If the bulk of the gas is of moderate density, then we can
simply adapt our estimate of $t_{\rm dis}$ for a spherically symmetric halo to 
the case of a far less massive clump. In this case, it is clear that the 
dissociation timescale will be short.
On the other hand, if the circumstellar gas is dense 
enough to form $\mHt$ via three body reactions, then 
$x_{\mHt} \simeq 1$, and the amount of $\mHt$ gas can be large. However, such
a high density also implies $f_{\rm abs} \simeq 1$, which goes some way to 
compensating for this. We find that in this case:
\begin{equation} 
t_{\rm dis} \simeq 20 M \left( \frac{N_{\rm dis}}{10^{48}} \right)^{-1} \:
\rm{yr}. 
\end{equation}
This is short compared to the stellar lifetime for any plausible value 
of the clump mass.

However, both of these cases are rather unrealistic. Not only do they neglect
the effects of ionizing radiation and stellar winds, which can be expected to 
help clear away any surrounding molecular gas, but they also assume spherical
symmetry. In reality, we would expect the gas surrounding newly formed 
primordial stars to be highly flattened. This is necessary for efficient 
outward transportation of angular momentum, in the assumed absence of 
magnetic fields. In this case, we would expect the circumstellar gas to 
survive for a much greater time, as it will receive less of the stellar flux. 
Nevertheless, it will not seriously affect our estimates as it will shield only
the small fraction of gas which lies in its shadow. 

\subsection{\hbox{H\,{\sc ii}} regions and stellar winds}
\label{hii}
Both ionizing radiation and stellar winds offer plausible alternative 
mechanisms by which $\mHt$ may be destroyed, and which may reduce the 
importance of $\mHt$ photodissociation. It is thus important to examine their
effects.

Stellar winds from low metallicity stars have recently been studied by 
Kudritzki~\shortcite{kud}. They are found to be several orders of magnitude 
less efficient 
than winds from solar metallicity stars, due to the lack of metal lines with 
which to drive them. It thus seems reasonable to neglect their effects.

Ionizing radiation, on the other hand, is rather more important. Ionizing flux
 from a massive star will lead to the formation of an \hbox{H\,{\sc ii}} 
region surrounding the star, whose size and subsequent evolution depend upon 
the density distribution near the star. A general treatment of the formation 
and evolution of this \hbox{H\,{\sc ii}} region requires solution of the 
coupled radiative transfer and hydrodynamics problem, which is clearly 
impractical. However, we can get some guidance as to the expected effects by 
examining some relevant cases where the high degree of symmetry allows for 
analytical solution.

The simplest case is that in which the gas surrounding the massive star is 
spherically symmetric and of uniform density. In this case, the 
growth of the \hbox{H\,{\sc ii}} region can be divided into two phases 
\cite{yor}.  
During the first phase, an R-type ionization front sweeps into the surrounding
 gas at supersonic velocity. As the radius of this front approaches the 
Str\"{o}mgren radius
\begin{equation} 
r_{\rm S} = \left( \frac{3 N_{\rm ion}}{4\pi n^{2} \alpha_{\rm B} }
\right)^{1/3}, 
\end{equation}
(where $N_{\rm ion}$ is the number of ionizing photons produced by the source 
per second and $\alpha_{\rm B}$ is the case B recombination coefficient), it 
slows and 
changes to a D-type front. As it does so, a shock front forms and detaches
from the ionization front, preceding it into the surrounding gas.
The subsequent expansion phase leads to the formation of a dense shell of 
\hbox{H\,{\sc i}}
gas between the shock and the ionization front. The thickness of this
 shell is small compared to the radius of the \hbox{H\,{\sc ii}} region. The 
latter is given approximately by \cite{yor}
\begin{equation} 
r_{\rm I} = r_{\rm S} \left[ 1 + \frac{7}{4} \frac{c_{\rm s}(t-t_{0})}
{r_{\rm S}} \right]^{4/7}. 
\end{equation} 
Here $c_{\rm s}$ is the isothermal sound speed in the \hbox{H\,{\sc ii}} 
region, and $t-t_{0}$ is the time since the formation of the shock. 
The expansion phase ends when the \hbox{H\,{\sc ii}} region attains pressure 
equilibrium with the surrounding gas. This occurs at a radius
\begin{equation} 
r_{\rm f} \simeq \left( \frac{2T_{2}}{T_{1}} \right)^{2/3} r_{\rm S}, 
\end{equation}
where $T_{1}$ and $T_{2}$ are the temperatures of the \hbox{H\,{\sc i}} and  
\hbox{H\,{\sc ii}} gas
respectively. For gas cooled by $\mHt$, the former will be approximately
 $200 \: \rm{K}$ (or smaller if HD cooling is important),  while the latter 
will be approximately $10^{4} \: \rm{K}$; thus $r_{\rm f} \simeq 20 r_{\rm S}$.
However, unless the gas density is extremely large, the main sequence lifetime
 of the central star will generally be too short to allow the 
\hbox{H\,{\sc ii}} region to 
reach this pressure equilibrium; in this case, the maximum size of the 
\hbox{H\,{\sc ii}} 
region is given by $r_{\rm I}$, evaluated at the end of the star's life.

If, rather than a uniform density distribution, we have an instead an 
inhomogeneous distribution, with uniform \emph{mean} density 
$\langle n \rangle$, then the above calculations still give us a reasonable
 estimate
for the size of the resulting \hbox{H\,{\sc ii}} region, as long as we 
replace $n^{2}$ by
$\langle n^{2} \rangle$. We generally express this by means of a clumping 
factor 
$C$ such that $C=\langle n^{2} \rangle / \langle n \rangle^{2}$. This results
 in a value
for $r_{\rm S}$ a factor $C^{1/3}$ smaller than in the uniform density case.

The other simple case we examine is the propagation of an \hbox{H\,{\sc ii}} 
region into
a radially decreasing density distribution. This problem has been studied by
Franco, Tenorio-Tagle \& Bodenheimer~\shortcite{ftb}, who show that if the 
density drops off more sharply than
$n \propto r^{-3/2}$  then the resulting \hbox{H\,{\sc ii}} region is 
unbounded by 
recombination. This implies that the entire halo may become ionized; however,
as this requires at least one ionizing photon per hydrogen atom (assuming
that secondary ionization by energetic photons is negligible), this 
cannot take place on a timescale less than 
\begin{eqnarray}
t_{\rm ion} & = & \frac{M}{\mu m_{\rm{H}} N_{\rm ion}} \nonumber \\ 
 & = & \frac{3.2 \times 10^{49} M}{N_{\rm ion}} \: \rm{yr}.
\end{eqnarray} 
where $N_{\rm ion}$ is the number of ionizing photons produced per second and
$M$ is the mass of gas in the halo.

To apply these calculations to our model halos, we need to specify 
$N_{\rm ion}$.
We can calculate this in a similar fashion to $N_{\rm dis}$, by modelling 
the star as a black body. Whilst this will inevitably be inaccurate,
the values we obtain compare well with those from the more detailed treatment
of Tumlinson \& Shull~\shortcite{shull+tum}. For the case of a 
$25 \, {\rm M}_{\odot}$ star considered
previously, we find that $N_{\rm ion} \simeq 7 \times 10^{48} \: \rm{s}^{-1}$.
 If we take 
the temperature of the \hbox{H\,{\sc ii}} region to be $10^{4} \: \rm{K}$, 
then 
\begin{equation} 
r_{\rm S} = 60 n^{-2/3} \: \rm{pc}, 
\end{equation}  
where we have used the case B recombination coefficient from 
Osterbrock~\shortcite{os}.

\begin{figure}
\centering
\epsfig{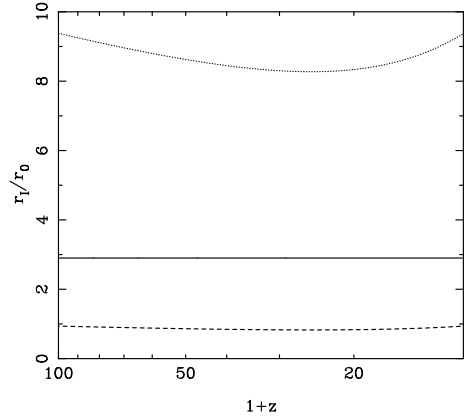}
\caption{The final size of the \hbox{H\,{\sc ii}} region formed by a 
$25 M_{\odot}$ star in halos of total mass $10^{5} M_{\odot}$ (dotted line) 
and $10^{8} M_{\odot}$ (dashed line) that is predicted if the density is 
assumed constant and equal to the central density of a truncated isothermal
sphere of the same mass, forming at the specified redshift. It is plotted in
units of $r_{0}$, the core radius of the truncated isothermal sphere. Also 
plotted is the radius at which the density profile of the sphere begins to
fall off more steeply than $r^{-3/2}$ (solid line). }
\label{ratio}
\end{figure}

For our uniform density halo model, this implies
\begin{equation} 
r_{\rm S} = \frac{8.1 \times 10^{4}}{(1+z)^{2}} \: \rm{pc}. 
\end{equation}
However, given the unrealistic nature of this density profile, this figure is 
probably not relevant. For the more realistic truncated isothermal sphere
model, we have
\begin{equation} 
r_{\rm S} = \frac{3.8 \times 10^{3}}{(1+z)^{2}} \: \rm{pc}, 
\end{equation}
assuming that the \hbox{H\,{\sc ii}} region is confined to the core of the 
density profile.
From this we can calculate $r_{\rm I}$, given some value for $t-t_{0}$. We 
choose to adopt a value of 2 Myr, corresponding to a typical OB star 
lifetime.\footnote{This is an overestimate, but will be reasonably accurate 
so long as the initial R-type phase is much smaller than the subsequent 
D-type phase.} In figure~\ref{ratio} we plot $r_{\rm I}$ (in 
units of the core radius $r_{0}$) for halos of total mass 
$10^{5} \: {\rm M}_{\odot}$ 
and $10^{8} \, {\rm M}_{\odot}$, at a range of redshifts. Also plotted is the 
radius at which the truncated isothermal sphere density profile becomes steeper
than $r^{-3/2}$; this occurs at approximately $2.9 r_{0}$. 

For halos in which $r_{\rm I} > r_{0}$, our analysis, which assumes constant 
density, begins to break down; for $r_{\rm I} > 2.9 r_{0}$, 
it breaks down completely, and the \hbox{H\,{\sc ii}} region will in reality 
become unbounded.
We see from figure~\ref{ratio} that this is true for all but the most massive 
halos.
Taking this at face value, it suggests that, as well as photodissociating 
$\mHt$ throughout a halo, a single massive star will also ionize 
the entire halo. However, this prediction may owe much to the overly simplistic
nature of our halo model.
 In a real halo, several factors are likely to combine to reduce our 
estimated values of $r_{\rm I}$. Significant clumping is likely, and will 
increase the effective density and hence decrease $r_{\rm I}$. Cooling will 
also increase 
the core density above that predicted by the truncated isothermal sphere model.
Finally, non-thermal pressures due to infalling gas or turbulence may also
restrict the expansion of the \hbox{H\,{\sc ii}} region. 

Given these uncertainties, plus the rather ad hoc nature of the adopted 
density profile, the final size of the first \hbox{H\,{\sc ii}} regions is 
rather unclear.
On the one hand, they may be comparable in size to the halo; on the other, 
they may prove to more closely resemble the ultracompact \hbox{H\,{\sc ii}} 
regions seen in
some local molecular clouds. To decide between these scenarios will require 
a better understanding of the density and velocity structure of the first 
protogalaxies than we currently possess, a deficiency that high resolution
numerical simulations are beginning to address \cite{abn}.  

Finally, it should be noted that even if the final \hbox{H\,{\sc ii}} region is
large, this will significantly affect our conclusions only in the smallest 
halos. In larger halos, $t_{\rm dis} < t_{\rm ion}$, so cooling
is still regulated in the first instance by photodissociation.
Similarly, clumps dense enough to preserve their $\mHt$ until collapse
will in general have $t_{\rm ff} < t_{\rm ion}$, and thus will only be 
affected by the \hbox{H\,{\sc ii}} region if they lie  close to the massive 
star. 

\subsection{Dust}
Dust, if present, can affect our simple picture in three ways. Firstly,
$\mHt$ can form by direct association on grain surfaces.
The rate of $\mHt$ formation by this process is independent of ionization. 
Consequently, at low temperatures the argument of Nishi \& Susa~\shortcite{ns}
 is no longer 
valid, and the $\mHt$ abundance may be significantly larger than $10^{-3}$.

Second, dust absorbs strongly in the Lyman-Werner bands, and will 
reduce the number of photons able to cause photodissociation.

Finally, dust implies the presence of metals, which may replace $\mHt$ as the
dominant coolant at low temperatures. This means that
 $\mHt$ dissociation need not imply
cessation of star formation, if the abundance of metals and dust is high enough
 to cool the gas.

With each possible effect, we associate a critical metal abundance, above which
the effect becomes important. We estimate these below.

\subsubsection{$\mHt$ formation on grains}
$\mHt$ forms on grain surfaces in the local ISM at a rate \cite{db}
\begin{equation} 
R_{\rm dust} = 6.0 \times 10^{-18} T^{1/2} n_{\rm H}^{2} \:
\rm{cm}^{-3} \: \rm{s}^{-1}, 
\end{equation}
where $n_{\rm H}$ is the \hbox{H\,{\sc i}} number density. 
If we assume both that dust in low metallicity gas has the same distribution
of grain sizes as that in the Milky Way, and that the ratio of dust to metals 
is independent of metallicity, then the rate of $\mHt$ formation on grains
 in metal-poor gas is simply
\begin{equation} 
R_{\rm dust} = 6.0 \times 10^{-18} T^{1/2} \left( 
\frac{Z}{Z_{\odot}} \right) n_{\rm H}^{2} \: \rm{cm}^{-3} \: \rm{s}^{-1}. 
\end{equation}
Both of these assumptions are somewhat \emph{ad hoc} but they are the simplest
that we can make, and should give us a reasonable idea of the importance of 
dust.

If destruction of $\Hm$ by means other than $\mHt$ formation is negligible,
as will be the case prior to star formation, then we can directly compare
 $R_{\rm dust}$ with the rate of the $\Hm$ gas phase reaction 
(equation~\ref{hmrate}). We see that at low metallicities the gas phase 
reaction will dominate, while at higher metallicities
formation on grain surfaces dominates. A reasonable definition of the critical
metallicity is that at which the two rates are equal. This is given by
\begin{equation} 
Z_{\rm crit} = \frac{T^{1/2} x_{\rm e}}{6} Z_{\odot}. 
\end{equation}
For a protogalaxy with temperature 
 $T=10^{3} \: \rm{K}$ and ionization 
$x_{\rm e}=3 \times 10^{-4}$ (corresponding to the residual abundance from 
recombination), this implies that
\begin{equation} 
Z_{\rm crit} = 1.6 \times 10^{-3} Z_{\odot}. 
\end{equation}
Thus, $\mHt$ formation on dust may become important at metallicities as small
as a few thousandths of the solar value.

Once stars have formed, the $\Hm$ pathway becomes far less effective as much
of it is now photodissociated rather than forming $\mHt$, while  $R_{\rm dust}$
is unaffected. However, by this stage the bulk of the $\mHt$ has already formed
and the `extra' $\mHt$ forming on dust for $Z \sim Z_{\rm crit}$ will increase
$t_{\rm dis}$ only slightly.

\subsubsection{Extinction}
To assess the importance of dust extinction, we calculate the optical depth 
due to dust at $1000 {\rm \AA}$. If we again assume that primordial dust is 
similar to that in the Milky Way, and that the dust to metal ratio is constant,
then the optical depth is given by
\begin{equation} 
\tau_{\rm d} = \sigma_{{\rm d},1000} N_{\rm H} \left( \frac{Z}{Z_{\odot}} 
\right), 
\end{equation}
where $\sigma_{{\rm d},1000}$ is the absorption cross-section at 
$1000 {\rm \AA}$, $N_{\rm H}$ the column density of hydrogen, 
again assuming a constant dust to metal ratio. Setting $\tau_{\rm d} = 1$, 
and adopting a value of $2 \times 10^{-21}\: \rm{cm}^{2}$ for 
$\sigma_{{\rm d},1000}$ \cite{db}, we can solve for the critical
metallicity at which the dust becomes optically thick. We find that
\begin{equation} 
Z_{\rm{crit}} = 5 \times 10^{20} N_{\rm H}^{-1} Z_{\odot}. 
\end{equation}

The critical metallicity thus depends upon the column density, and consequently
upon the properties of the individual halo or clump. The range of column 
densities included in our model is large, being approximately $10^{19}$ to 
$10^{23} \: \rm{cm}^{-2}$ for halos and $10^{21}$ to $10^{25} \:
\rm{cm}^{-2}$ for clumps. Adopting the median values of $10^{21} \:
\rm{cm}^{-2}$ and $10^{23} \: \rm{cm}^{-2}$ as representative of `typical'
halos and clumps, this implies critical metallicities of $0.5 \: Z_{\odot}$ and
$5 \times 10^{-3} \: Z_{\odot}$ respectively. This suggests that dust 
absorption in halos is unlikely to be important until substantial enrichment 
has occured, while clumps become shielded at much smaller metallicities.

\subsubsection{Alternative coolants}
The presence of metals makes available a large number of alternative cooling
processes, such as fine structure emission from \hbox{O\,{\sc i}} and 
\hbox{C\,{\sc ii}}, or rotational emission from CO. At a sufficiently
high metallicity, these dominate the low temperature cooling. To properly
assess the metallicity at which this occurs is a complex problem, lying
far beyond the scope of this paper. However, a study along these lines has
recently been performed by Omukai~\shortcite{om}. He finds that cooling from 
sources
other than $\mHt$ typically becomes dominant at a critical metallicity
$Z_{\rm crit} \simeq 10^{-2} Z_{\odot}$. While the situation he considers --
protostellar collapse to very high density, with no UV radiation field -- 
differs markedly from our own, the differences are unlikely to make
alternative coolants significantly more effective. 
 
\section{Conclusions}
In this paper we have examined the destruction of $\mHt$ within primordial 
protogalaxies by ultraviolet radiation from the first massive stars to 
form within them. To do this we have constructed a simple analytical model of 
photodissociation which incorporates the essential physics and which should
give a reasonable order of magnitude estimate of the time taken to destroy 
the bulk of the $\mHt$.
Using this model, we have shown that for diffuse gas in small protogalaxies 
the conclusions arrived at by Omukai \& Nishi are broadly correct:
 a single massive star will produce sufficient ultraviolet
 radiation to suppress $\mHt$ cooling within its own lifetime. On the other 
hand, in larger protogalaxies the photodissociation timescale becomes 
comparable to the stellar lifetime, and the effects on gas cooling are 
correspondingly smaller.

We have also examined the effects of photodissociation upon dense clumps of 
gas, such as we expect to find in star forming regions. Using a  simplified 
clump model, we find that there is a density-dependent critical distance, 
$D_{\rm crit}$,
beyond which the photodissociation timescale is longer than the clump 
free-fall timescale. For moderate density clumps, $D_{\rm crit}$ is 
comparable to the dimensions of the star-forming region, but it decreases 
sharply as clump density increases. 
Since clumps can continue to collapse only if cooled by $\mHt$, it
is reasonable to assume that the survival of $\mHt$ is a prerequisite for star
formation. Thus, stars will form only in clumps that lie sufficiently far
from any massive star. An alternative way of considering this is to suppose 
that stars will only form in clumps above some density threshold, where this 
is chosen such that the corresponding $D_{\rm crit}$ is much smaller than the 
dimensions of the protogalaxy. In this case, an appropriate value for the 
density threshold is approximately $10^{6} \: \rm{cm}^{-3}$, although
this is at best a rough estimate.
 
We have also examined a number of factors which may complicate this simple
 picture. The most important prove to be ionizing radiation and the presence 
of dust and metals. 

The effects of ionizing radiation are sensitive to the 
small-scale density distribution within the halo, and are thus hard to assess
accurately. However, we have shown that even if there is sufficient flux to
ionize most or all of the halo, this will generally take longer 
than $\mHt$ photodissociation. Consequently, it should have little 
effect on our central results.

Dust and metals can lead to more significant changes in the physics, but these
begin to take effect only once the metallicity of the gas exceeds a few 
thousandths of the solar value.

Our overall picture of star formation in the first protogalaxies is one 
in which the formation of dense clumps plays a central role. Ultraviolet 
feedback is effective at suppressing cooling in diffuse gas, but has little 
effect on dense clumps. Moreover, while it is 
likely that supernovae will eventually remove all of the gas from the system, 
we expect this to occur \emph{after} clump collapse. An
upper limit to the star formation efficiency is thus set by the fraction of 
gas that has formed into dense clumps by the time that the first massive stars
ignite. This may be small, but it is too early to conclude this with any
real certainty, and thus scope for significant star formation 
within the first cosmological objects still exists. 

\section*{Acknowledgements}
We are grateful to Bruce Draine for correcting an error in the calculation
of dissociation rate, and for useful comments from the referee, John
Black. SCOG acknowledges support by a PPARC studentship.

\end{document}